`
%
%
%
%
%
%
%
%
%
%
%
%
%
%
\input phyzzx
%
%
\catcode`\@=11 
\def\papersize{\hsize=40pc \vsize=53pc \hoffset=0pc \voffset=-2pc
   \advance\hoffset by\HOFFSET \advance\voffset by\VOFFSET
   \pagebottomfiller=0pc
   \skip\footins=\bigskipamount \normalspace }
\catcode`\@=12 
\papers
\vsize=23.cm
\hsize=15.cm

\Pubnum={LPTENS-98/18 \cr
{\tt hep-th@xxx/9805070} \cr
May 1998}

\date={}
\pubtype={}
\titlepage
\title{{\bf DLCQ of M-theory as the light-like limit}
}
\author{Adel~Bilal
\foot{Partially supported by the
European Commision under TMR contract FMRX-CT96-0090.}
}
\vskip .5cm
\address{
CNRS - Laboratoire de Physique Th\'eorique de l'\'Ecole
Normale Sup\'erieure
\foot{{\rm unit\'e propre du CNRS, associ\'ee \`a l'\'Ecole Normale
Sup\'erieure et l'Universit\'e Paris-Sud}}    \break
24 rue Lhomond, 75231 Paris Cedex 05, France  \break
{\tt bilal@physique.ens.fr} \break
and  \break
Institut de Physique, Universit\'e de Neuch\^atel  \break
rue Breguet 1, 2000 Neuch\^atel, Switzerland
\foot{{\rm address after September 1, 1998 }}
}

\vskip 0.5cm
\abstract{ 
We investigate whether DLCQ of M-theory can be defined as a limit of 
M-theory compactified on an almost light-like circle. This is of particular 
interest since the proofs of the matrix description of M-theory by Seiberg 
and Sen rely on this assumption. By the standard relation between 
M-theory on $S^1$ and IIA string theory, we translate this question into 
the corresponding one about the existence of the light-like limit of IIA 
superstring theory for {\it any} string coupling $g_s$. 

We  argue that perturbative string loop amplitudes should have a finite 
and well-defined light-like limit provided the external momenta are 
chosen to correspond to a well-defined DLCQ set-up. On the 
non-perturbative side we consider states and amplitudes. We show that 
an appropriate class of non-perturbative states (wrapped D-branes) 
precisely have the right light-like limit. We give some indications that 
non-perturbative corrections to string amplitudes, too, may behave as 
required in the light-like limit. Having perturbative and non-perturbative 
evidence, this suggests that type IIA superstring theory as a whole has 
a well-defined light-like limit (for any string coupling $g_s$) and hence 
that the same is true for M-theory.
}

%

\endpage
\pagenumber=1

\def\PL #1 #2 #3 {Phys.~Lett.~{\bf #1} (#2) #3}
\def\NP #1 #2 #3 {Nucl.~Phys.~{\bf #1} (#2) #3}
\def\PR #1 #2 #3 {Phys.~Rev.~{\bf #1} (#2) #3}
\def\PRL #1 #2 #3 {Phys.~Rev.~Lett.~{\bf #1} (#2) #3}
\def\CMP #1 #2 #3 {Comm.~Math.~Phys.~{\bf #1} (#2) #3}

\def\b{\beta}

\def\m{\mu}
\def\n{\nu}
\def\nsr{\nu_{sr}}

\def\re{R_{11}}

\def\rs{R_s}
\def\e{\epsilon}

\def\t{\tau}

\def\ap{\alpha'}
\def\rap{\sqrt{\ap}}

\def\to{\rightarrow}
\def\Im{{\rm Im}}

\def\kcl{K_{\rm cl}}

\REF\WIT{E. Witten, {\it String theory dynamics in various
dimensions}, \NP B443 1995 85 , {\tt hep-th/9503124}.}

\REF\SEI{N. Seiberg, {\it Why is the matrix model correct?}, 
Phys.Rev.Lett. {\bf 79} (1997) 3577, {\tt hep-th/9710009}.}

\REF\SEN{A. Sen,  {\it D0-branes on $T^n$ and matrix theory}, {\tt hep-th/9709220}.}

\REF\ABM{A. Bilal, {\it M(atrix) theory : a pedagogical introduction}, {\tt hep-th/9710136}.}

\REF\SUSS{L. Susskind,  {\it Another conjecture about M(atrix) theory}, {\tt hep-th/9704080}.}

\REF\BFSS{T. Banks, W. Fischler, S.H. Shenker and L. Susskind, {\it
M theory as a matrix model: a conjecture}, \PR D55 1997 5112 , 
{\tt hep-th/9610043}.}

\REF\PH{S. Hellerman and J. Polchinski, {\it Compactification in the lightlike limit}, 
 {\tt hep-th/9711037}.}

\REF\ABD{A. Bilal, {\it A comment on compactification of M-theory on an (almost) light-like circle},
to appear in Nucl. Phys. B, {\tt hep-th/9801047}.}

\REF\GUIJ{A. G\"uijosa, {\it Is physics in the infinite momentum frame independent of the compactification
radius?}, {\tt hep-th/9804034}.}

\REF\KP{E. Kiritsis and B. Pioline, {\it On $R^4$ threshold corrections in type IIB string theory and
(p,q)-string instantons}, {\tt hep-th/9707018};\nextline
M.B. Green and M. Gutperle, {\it Effects of D-instantons}, \PL B398 1997 69 , {\tt  hep-th/9701093}, 
and {\it D-particle bound states and the D-instanton measure}, {\tt hep-th/9711107}.}

\REF\PIO{B. Pioline, {\it D-effects in toroidally compactified type II string theory}, {\tt hep-th/9712155};
\nextline
M.B. Green and P. Vanhove, {\it D-instantons, strings and M-theory},  {\tt hep-th/9704145}. }

\REF\GSW{M. Green, J. Schwarz and E. Witten, {\it Superstring theory}, vol. 2, Cambridge University
Press, 1987.}

\REF\WATI{W. Taylor, {\it D-brane field theory on compact spaces}, \PL B394 1997 283 ,  
{\tt hep-th/9611042}. }


{\bf \chapter{Introduction and summary}}

The notion of eleven-dimensional M-theory emerged [\WIT] as the strong-coupling limit of IIA
superstring theory. More precisely, M-theory compactified on a (space-like) circle of radius $\re$ is
identical to IIA superstring theory with coupling $g_s=\re/\rap$. While this sounds like a mere
definition, the non-trivial conjecture is that in the $g_s\to \infty$ limit, the resulting theory has
eleven-dimensional Lorenz invariance. Once one accepts this conjecture, a statement about
M-theory can be translated into a statement about IIA
superstring theory at generic coupling. More precisely, any question one might ask in M-theory
compactified on a manifold $S^1\times K$ can be translated into, and in principle be answered within
IIA theory on $K$, provided we keep the string coupling $g_s$ {\it generic} and do not restrict
ourselves to perturbative physics (in $g_s$) only. By taking the appropriate $g_s\to\infty$ limit, IIA
on $K$ of course corresponds to M on $K$.

The question we want to ask here is the following: Does M-theory compactified on a space-like circle
of radius $\rs$ has a well-defined limit as $\rs\to 0$. This is of quite some interest since it is the basic
assumption in Seiberg's [\SEI] and Sen's [\SEN] proof that the DLCQ of M-theory is given by the
finite $N$ matrix model (for a pedagogical review, see [\ABM]). Let me recall that Susskind
conjectured [\SUSS] that the quantization of M-theory compactified on a light-like circle of radius $R$
(discrete light-cone quantization = DLCQ) in a sector of fixed $p_-={N\over R}$ is given by a ${\rm
U}(N)$ matrix quantum mechanics as obtained by reduction from ten-dimensional super Yang-Mills.
The advantage of this DLCQ of M-theory with respect to the infinite momentum frame of Banks et al
[\BFSS] is that various dualities are already manifest at finite $N$. Seiberg and Sen consider this
DLCQ of M-theory as the limit of a compactification on an almost light-like circle, which in turn is
Lorenz equivalent to compactification on a space-like circle of radius $\rs$. Since $2\pi\rs$ is the
proper (Lorenz invariant) length of all the circles, the light-like limit is recovered for $\rs\to 0$. Under
the assumption that this $\rs\to 0$ limit of M-theory makes sense, Seiberg and Sen argued, using
various dualities, that the DLCQ of M-theory is indeed equivalent to IIA theory in a particular limit
where $g_s\to 0$ and the string mass diverges, so that all that is left in a sector of non-zero D0
brane charge $\sim N$ is the corresponding ${\rm U}(N)$ matrix model describing the dynamics of the
open string ground states.

The key question remaining open is whether this $\rs\to 0$ limit exists. In case it does, it should actually
define what is meant by DLCQ of M-theory. To elucidate this issue, Hellerman and Polchinski [\PH]
considered  quantum field theories compactified on a circle, in the $\rs\to 0$ limit. They found that, typically,
one-loop diagrams with vanishing $p_-$ transfer are plagued with divergences $\sim {1\over \rs}$, thus
casting serious doubt on the existence of the required limit in M-theory. However, M-theory is certainly
very different from ordinary quantum field theory, in particular due to the existence of extended objects: membranes
and five-branes. In this respect it much more ressembles string theory where the existence of the winding
modes of the string plays a crucial role. The present author has investigated this $\rs\to 0$ limit for a
four-point one-loop amplitude in type II superstring theory [\ABD] and found that this limit is perfectly
well-defined, even for  vanishing $p_-$ transfer (the potentially troublesome case). It was also shown
[\ABD] that this limit reproduces the result of a direct DLCQ computation of the string amplitude.

While the point of view in [\ABD] was to consider superstring theory as a possible analogy with M-theory,
here we take a different attitude. We consider M-theory on a circle $S^1$ of radius $\re$ as IIA
theory with $g_s=\re/\rap$. The statement we would like to prove is:

\noindent
{\bf statement A}: ``The IIA superstring theory with coupling $g_s$, compactified on a space-like circle
$S^1$ of radius $\rs$
has a well-defined limit as $\rs\to 0$."

\noindent
As discussed above, this then is equivalent to the 

\noindent
{\bf statement B}: ``M-theory compactified on (space-like) $S^1\times S^1$ with radii $\re=\rap g_s$ and $\rs$ has a
well-defined limit as $\rs\to 0$."

\noindent
If statement A can be proved (as uniform convergence) for {\it any} string coupling $g_s$, then we have
shown the 

\noindent
{\bf statement C}: ``M-theory compactified on a space-like circle of radius $\rs$ has a well-defined limit as
$\rs\to 0$."

\noindent
This then would fill the gap in Seiberg's and Sen's proofs.

Although a complete and rigorous proof of statement A for any $g_s$ certainly is beyond reach, we
nevertheless now have much perturbative and non-perturbative knowledge about the IIA theory at our
disposal in order to check this statement to quite some extent. This is the purpose of the present note.

On the perturbative side one has to consider arbitrary $N$-point genus-$g$ amplitudes in type II
superstring theory compactified on a circle of radius $\rs$ and check that the $\rs\to 0$ limit is
well-behaved. As explained in [\ABD], one has to consider amplitudes for external states with non-vanishing
momenta in the compact direction (see [\GUIJ] for a discussion of non-vanishing winding numbers also). In
[\ABD] the simplest case, $N=4$, $g=1$, was studied in detail. Already there it was clear that the same
argument should apply to any $N$-point genus-1 amplitude. Here we will argue that similarly the $\rs\to 0$
limit should also exist for any genus-$g$ amplitude.

On the non-perturbative side, the first think to do is to simply look at the spectrum of BPS states. We will
argue that precisely those D-brane configurations that had a finite mass before compactifying the
space-like direction do scale appropriately to survive and make sense in the light-like
($\rs\to 0$) limit. More ambitiously, one can look at non-perturbative corrections to $N$-point scattering
amplitudes. Some important information can already be extracted from the D-instanton corrections to the $R^4$
coupling of the low-energy effective action [\KP]. We will see that these corrections do depend on $\rs$ in
exactly the way needed so that   a well-defined  $\rs\to 0$ limit of the full amplitude might exist!

Thus although not a proof, I believe that the perturbative and non-perturbative evidence presented in this
note is rather encouragingly pointing towards the existence of a well-defined light-like limit ($\rs\to 0$) of
M-theory.

{\bf \chapter{Kinematics}}

Let me begin by briefly describing the limit we are interested in. We want to study type II
superstring theory compactified on a space-like circle of radius $\rs$ in the limit $\rs\to 0$. This is Lorenz
equivalent to a compactification on an almost light-like circle, with $\rs\to 0$ corresponding to the light-like
limit [\SEI,\PH,\ABD]. We will take the space-like compactified direction to be $x^9$ ($\rs\equiv R_9$), 
so that the
corresponding momenta $p_9$ are quantized:
$$x^9\simeq x^9 + 2\pi \rs \quad , \quad p_9={n\over \rs} \ .
\eqn\di$$
Since we are interested in the $\rs\to 0$ limit it is convenent to write $\rs = \e R_0$ where $R_0$ is kept
fixed. Using a boost with parameter $\b=(1-\e^2/2)/(1+\e^2/2)$ we get  a Lorenz equivalent coordinate
system (denoted $\tilde x^\m$  with $\tilde x^\pm =(\tilde x^0 \pm \tilde x^9)/\sqrt{2}$) where
$$ \tilde x^+ \simeq \tilde x^+ + \pi \e^2 R_0 \quad , \quad \tilde x^- \simeq \tilde x^- - 2\pi R_0 \ .
\eqn\dii$$
For $\e\to 0$ this gives a light-like compactification. 
Let's make this more precise. Introduce yet another coordinate system by $\hat x^-=\tilde x^-, \ \hat t
=\tilde x^+ +\e^2 \tilde x^- /2$. Then $\hat t$ is a standard non-compact coordinate, while $\hat x^-$ still
has period $2\pi R_0$. The metric is ${\rm d}s^2=-{\rm d} \hat t {\rm d}\hat x^- + \e^2 ({\rm d} x^-)^2$ so
that the light-like limit is indeed $\e\to 0$. The momentum $\hat p_-$ is quantized as  $\hat p_-={n\over
R_0}$ while $\hat p_t$ takes continuous values. The relation between the momenta in the different frames
is easily seen to be
$$\eqalign{
\tilde p_+ = \hat p_t \quad &, \quad \tilde p_- = \hat p_- + {\e^2\over 2} \hat p_t \ , \cr
p_9={1\over \e} \hat p_-  \quad &, \quad  p_0={1\over \e} \hat p_- + \e \hat p_t \ . \cr
}
\eqn\kin$$
Now $\hat p_t =\tilde p_+$ is the DLCQ hamiltonian and should have a finite limit as $\e\to 0$. Then,
since $\hat p_-={n\over  R_0}$, one has $p_9={n\over \e R_0}$ so that the space-like momentum blows
up . Also, $p_0 = {n\over \e R_0} + {\cal O}(\e)$ so that the energies in this frame also scale as $1/\e$.

{\bf \chapter{Perturbative evidence}}

We now want to study string scattering amplitudes compactified on a space-like circle of radius
$R_s\equiv R_9 =\e R_0$. We have just seen that the 
external momenta we are interested in have fixed, generically non-zero $\hat p_-$, i.e. fixed
non-zero $n$. This means that we want to look at spacelike momenta in the compact 
direction of the form $p_9={n\over R_s}={n\over \e R_0}$ that blow up as $\e\to 0$.

\section{Genus one}

Let me first briefly recall the proof of [\ABD] that the four-point one-loop amplitude of type II
superstring theory compactified on a space-like circle has a well-defined limit as $\rs\to 0$.  This amplitude
for massless (in the ten-dimensional sense) external states with fixed momentum quantum numbers $n_r,\
r=1, \ldots 4$ was given in [\ABD] to be
$$\eqalign{
A_{\rm cl}^{(4)}=&{(\pi \kappa)^4 \over \ap^5} \kcl 
\int {{\rm d}^2\t\over (\Im\t)^2}\,  \prod_{r=1}^3 {{\rm d}^2\n_r\over \Im\t}\
\prod_{s>r} \chi(\nsr,\t)^{\ap k_r\cdot k_s}  \cr
&\times \sum_{n,m} 
{\ap\over \rs^2} \exp\left\{ -\pi {\ap \over \rs^2}
{1\over \Im\t} \left\vert m+n\t +\sum_{s=1}^4 n_s\n_s \right\vert^2 \right\} \cr}
\eqn\diii$$
where $\chi(\n,\t)=2\pi \exp[-\pi (\Im\n)^2/\Im\t]\vert \theta_1(\n,\t)/\theta_1'(\n,\t)\vert$,
and $k_r\cdot k_s$ denotes the full
ten-dimensional scalar product of the external momenta (we write $k_r$ rather than $p_r$, as
customary),
while $\kcl $ is the standard kinematic factor already present in the closed string tree amplitude
[\GSW] .
The important point that was noticed in [\ABD] is that although for vanishing $n_r$ the amplitude \diii\ diverges
as $\rs\to 0$, for at least one non-vanishing $n_r$ it has a finite limit. Indeed, let $\rs=\e R_0$ so that the
relevant $\e$-dependent factor is
$${1\over \e^2} \left( {\ap\over R_0^2}\right) \exp\left\{ -{\pi\over \e^2}  \left( {\ap\over R_0^2}\right) 
{1\over \Im\t}  \left\vert m+n\t +\sum_{s=1}^4 n_s\n_s \right\vert^2 \right\} 
\eqn\div$$
(in ref. [\ABD] we took $R_0^2=\ap$). As $\e\to 0$ this yields a complex delta function:
$$\div \quad \to \quad   \Im\t\ \   \delta^{(2)} \left(  m+n\t +\sum_{s=1}^4 n_s\n_s \right) \ .
\eqn\dv$$
The net effect of the sum over $n$ and $m$ and the integration over the moduli $\n_s$ then is to replace
one of the $\n_s$-integrations, say the $\n_3$-integration, by a discrete sum over a lattice of $n_3^2$
values on the world-sheet torus.

It was shown in [\ABD] that this amplitude exactly has the singularities required by unitarity, and no more.
In particular the case $n_1=-n_2,\ n_3=-n_4$ of vanishing momentum transfer in the compact direction
(which was the dangerous case [\PH]) is perfectly finite, except for poles corresponding to on-shell
intermediate states, as it should.

It was already clear in [\ABD] that the restriction to only four external states (amplitudes with less
than four external states vanish)
was of not much relevance for our argument. The basic point was the exponential factor, coming
from the zero-modes (momenta and winding modes) after a partial Poisson resummation, and the ${1\over
\rs^2}$ factor from the measure of the momentum modes and again the partial Poisson resummation. It is
pretty clear that for $N>4$ external states, the expression \div\ would simply be replaced by
$${1\over \e^2} \left( {\ap\over R_0^2}\right) \exp\left\{ -{\pi\over \e^2}  \left( {\ap\over R_0^2}\right) 
{1\over \Im\t}  \left\vert m+n\t +\sum_{s=1}^N n_s\n_s \right\vert^2 \right\} \ .
\eqn\dvi$$
In the $\e\to 0$ limit this again leads to the complex delta function (with the sum over $s$ now
running from 1 to $N$) with the same
net effect of discretizing one of the moduli $\n_s$. One obtains a finite amplitude having only the
singularities required by unitarity.

\section{Higher genus}

For the one-loop amplitude, the factor of $1/\e^2 \sim 1/\rs^2$ was to be expected from T-duality. Indeed,
as compared to the tree-level  amplitude, the one-loop amplitude carries an extra factor of $g_s^2={\rm
e}^{2\phi}$. Under T-duality, ${\rm e}^{2\phi}/\rs^2$ converts to ${\rm e}^{2\tilde\phi}$ with no {\it explicit}
$\rs$ dependence as the dual radius $\tilde\rs=\ap/\rs$ goes to $\infty$. The same type of argument shows
that for a genus-$g$ amplitude there must be a factor $\rs^{-2g}$. Indeed, we get a factor of $\rs^{-2}$ for
{\it each} handle from the momentum measures and partial Poisson resumming of the winding modes. This
factor $\rs^{-2g}\sim \e^{-2g}$ must combine with appropriate exponentials $\prod_{i=1}^g \exp\left( -
(\pi/\e^2) \vert \ldots \vert^2 \right)$ to give a product of $g$ delta functions. These delta functions then
e.g. fix one of the insertion points of external states and $g-1$ of the Teichm\"uller parameters describing
the genus-$g$ Riemann surface. 

Rather than working this out in general, let me only consider a genus-2 surface in the limit where it looks
like two tori joined by a long and narrow tube. An $N=N_1+N_2$-point amplitude then looks like the product
of an $N_1+1$-point one-loop amplitude with an $N_2+1$-point one-loop amplitude integrated over the
modular parameter describing the long narrow tube. In the $\rs\to 0$ limit, each one-loop amplitude then
indeed gives a complex delta function as discussed in the previous subsection, and the net effect is to
discretize e.g. the insertion point of the long narrow tube on one of the tori, as well as  the insertion
points of one of the external states. According to our discussion for $g=1$, this amplitude must be finite
and only have those singularities that are required by unitarity.
Although we have only discussed this very special geometry of the two-loop amplitude, it is already quite
encouraging and shows how a similar result may well hold for the general case.

{\bf \chapter{Non-perturbative evidence}}

\sectionnumber=0

\section{D-branes}

In section 2 we have argued that we should be interested in states with energies $p_0$ scaling as $1/\e$
because only such states can correspond to a fixed non-vanishing $\hat p_-$ and finite DLCQ energy.
We will now show that this is indeed satisfied for all D-branes that had finite energy before
compactification of $x^9$.

First consider D0 branes. Before compactifying $x^9$, a D0 brane has a finite mass $T_0$. Upon
compactification of $x^9$ on a circle of radius $R_9=\e R_0$, the D0  becomes a D1 wrapped around the {\it
dual} circle of radius ${\ap\over R_9}$ [\WATI]. It thus has an energy $T_1 2\pi  {\ap\over R_9}
={2\pi T_1 \ap\over \e R_0}$ which indeed scales as $1/\e$ as required.

Next look at a D2 brane. To get a finite mass/energy in the theory with non-compact $x^9$ we can
compactify two other (transverse) directions, say $x^7$ and $x^8$ with radii $R_7$ and $R_8$ and
wrap the D2 around this $T^2$. If we now compactify $x^9$ with radius $R_9=\e R_0$, the D2 becomes
a D3 wrapped around a $T^3$ with radii $R_7, R_8$ and ${\ap\over R_9}$ so that its energy is 
$T_3 (2\pi )^3 R_7 R_8 {\ap\over R_0 \e}$, 
again $\sim 1/\e$, as claimed. However, one may also start with a D2 in the 8-9
direction (wrapped or not in the 8 direction). If $x^9$ is not compact this D2 has  infinite extent, hence
infinite energy. Upon compactifying $x^9$ it  then becomes a D1 in the 8 direction. There is no way it's
energy can scale as $1/\e$. The argument is the same for higher branes.

Thus we see that D-branes that had finite energy before compactifying $x^9$, have an energy scaling
as $1/\e$ once $x^9$ is compactified with radius $R_9=\e R_0$, showing  they have finite DLCQ energy.
Hence the non-perturbative states made up from D-branes of finite energy behave in just the right way
to survive and make sense in the light-like limit.

\section{Non-perturbative amplitudes}

Of course, it is a formidable task to work out non-perturbative corrections to an arbitrary superstring
amplitude in general. However, there are some limiting cases of such amplitudes where the full series of
non-pertubative corrections is known. We may look at the low-energy limit where one can extract various
terms of the effective action from the full string amplitudes. Particularly interesting are the $R^4$-couplings
[\KP] because they are BPS protected and can only receive contributions from string tree-level, one-loop and
non-perturbative effects. Let us first discuss in general what one might learn about
the light-like limit from these couplings  and then consider the explicit results of [\KP,\PIO].

A priori, one might expect that one cannot extract any useful information for our purpose from a {\it
low-energy} effective action, because, as we saw above, we are interested in scattering amplitudes with
momenta in the compact dimension and energies diverging as $1/\e$ in the light-like limit. Nevertheless, the
low-energy effective couplings or amplitudes being the low-energy {\it limits} of some corresponding
expressions valid at any energy scale, the former contain valuable information about the possible forms of
the latter. 

To illustrate this point, consider again the four-point one-loop amplitude \diii\ of section 3 {\it before} taking
the $\rs\equiv R_9\to 0$ limit. Its low-energy limit (i.e. for $p_9^r=0$) gives, among others, the
one-loop contribution to the $R^4$-coupling of the low-energy effective action. In particular, the dependence
on the compactification radius is ${\ap\over R_9^2}$. This is the well-known reason why in the low-energy
limit one must combine $R_9\to 0$ with $\ap\to 0$. But we can turn the argument around. As we have seen in
section 3, this ${1\over R_9^2}$-dependence of the (one-loop) low energy effective $R^4$-coupling is a
necessary condition for the $R_9\to 0$ limit of the high-energy ($p_9^r\sim {1\over \e}$) amplitude to exist.
The key point was to combine the ${1\over R_9^2}$ with the exponential zero-mode factor into \div\ which
gave the delta function \dv.

Let us now turn to non-perturbative (D-instanton) corrections to such a string scattering amplitude.
Because we are looking at the corrections to the same process as in section 3, we would expect 
a similar exponential zero-mode factor to be present again. Rather than depending on the torus modular
parameter $\tau$ as in \div\ it
might depend on the moduli of the D-instantons. We do not know these zero-mode factors and,
of course, the
present discussion is highly speculative. It nevertheless seems a fair guess that again one has a product of
${1\over R_9^2}$ and some other  factor and that both combine into some delta function as $R_9\to 0$ so
that this limit turns out to be finite. At low energies however, this other factor should become trivial and only the
${1\over R_9^2}$ should remain. 
Although this is not the only possible scenario,\foot{
Obviously, one could also expect other scenarios leading e.g. to multiple delta functions as we have
seen for the perturbative genus $g>1$ amplitudes in section 3.2. Then one could e.g. have a factor
$R_9^{-2N}$ for D-instantons of charge $N$.
}
it is the simplest one and it seems reasonable to
expect that all the non-perturbative corrections to the
$R^4$-coupling in the effective action of the IIA string compactified on $S^1$ with radius $R_9$ shoud
behave as ${1\over R_9^2}$. 

The full series of these corrections is given in [\PIO] and reads (including also
tree-level and one-loop contributions)
$${1\over R_9} f^{\rm IIA}_{D=9} = 2 \zeta(3) {\rm e}^{-2\phi} + {2\pi^2 \ap\over 3 R_9^2} 
+{4\pi \rap\over R_9} {\rm e}^{-\phi} \sum_{m,n\ne 0} \left\vert {m\over n}\right\vert
K_1(2\pi \vert m n \vert R_9  {\rm e}^{-\phi} /\rap )  {\rm e}^{2\pi i m n {\cal A}} 
\eqn\qi$$
where $K_1$ is the Bessel function and ${\cal A}$ the Wilson line of the RR one-form on the circle. The
contribution of a D-instanton of charge $N=nm$ can be read off to be
$${4\pi \rap\over R_9} {\rm e}^{-\phi} \sum_{n\vert N} {\vert N \vert \over n^2} 
K_1(2\pi  \vert m n \vert R_9  {\rm e}^{-\phi}/\rap )
 {\rm e}^{2\pi i N {\cal A}} \ .
\eqn\qii$$
For $R_9\to 0$, the argument of this Bessel function becomes vanishingly small and, using $K_1(z) \sim
z^{-1} + {\cal O}(z, z\log z)$ as $z\to 0$, one gets for the D-instanton of charge $N$
$$ {2\over R_9^2}  \left( \sum_{n\vert N} {1\over n^2} \right)  {\rm e}^{2\pi i N {\cal A}} \sim {1\over R_9^2} \ .
\eqn\qiii$$
This is exactly as expected from our discussion! Although this is only rather indirect evidence that the
non-perturbative contributions to the full string scattering amplitudes with non-vanishing $p_9^r={n^r\over
R_9}$ have a finite $R_9\to 0$ limit, it nevertheless points in the right direction.
Note that \qiii\ is valid as long as $N R_9 \ll {\rm e}^\phi/\rap = g_s$. Thus the limit $R_9\to 0$ cannot be uniform
for all $N$ in the full expression \qi. Also, for $R_9\to 0$ we used the small $z$ asymptotics of $K_1(z)$,
while to make explicit the instanton order one woud have to use the large $z$ asymptotics 
$K_1(z) \sim \sqrt{\pi/ (2z)} {\rm e}^{-z}$ to obtain an expansion in powers of 
$\exp\left[-2\pi ((R_9/\rap) {\rm e}^{-\phi} + i {\cal A}) \right]$. Clearly, small $R_9$ and small ${\rm e}^\phi$ are
very different regimes. In particular, \qiii\ does not depend on $g_s={\rm e}^\phi$ at all, exactly as the
one-loop contribution.

{\bf \chapter{Conclusion}}

We have argued that perturbative string loop amplitudes should have a finite and well-defined light-like limit
provided the external momenta are chosen to correspond to a well-defined DLCQ set-up. On the
non-perturbative side we considered states and amplitudes. We showed that the appropriate class of
non-perturbative states (D-branes with finite energy for non-compact $x^9$) have precisely the right
light-like limit. We had less precise things to say about non-perturbative corrections to string amplitudes, but
still displayed some indications that they, too, might behave as required in the light-like limit. Having
perturbative and non-perturbative evidence, this suggests that type IIA superstring theory as a whole has
a well-defined light-like limit (for any string coupling $g_s$) and hence that the same is true for M-theory.

\refout

\end